\documentclass[twocolumn,prd]{revtex4}%
\usepackage{graphicx}
\usepackage{amsmath}
\usepackage{amsfonts}
\usepackage{amssymb}%
\providecommand{\U}[1]{\protect\rule{.1in}{.1in}}
\newcommand{\f}{\begin{equation}}
\newcommand{\ff}{\end{equation}}
\newcommand{\fa}{\begin{eqnarray}}
\newcommand{\ffa}{\end{eqnarray}}

\begin{document}
\title{Modified dispersion relations and black hole physics}
\author{Yi Ling$^{1,2}$}\email{yling@ncu.edu.cn}
\author{Bo Hu$^{1}$}\email{bohu@ncu.edu.cn}
\author{Xiang Li$^{1,2}$}\email{xiang.lee@163.com}
\affiliation{${}^1$ Center for Gravity and Relativity, Department
of Physics, Nanchang University, 330047, China}
\affiliation{%
${}^2$ CCAST (World Laboratory), P.O. Box 8730, Beijing
   100080, China}
\begin{abstract}
A modified formulation of energy-momentum relation is proposed in
the context of doubly special relativity. We investigate its
impact on black hole physics. It turns out that such modification
will give corrections to both the temperature and the entropy of
black holes. In particular this modified dispersion relation also
changes the picture of Hawking radiation greatly when the size of
black holes approaching the Planck scale. It can prevent black
holes from total evaporation, as a result providing a plausible
mechanism to treat the remnant of black holes as a candidate for
dark matter.
\end{abstract}

 \maketitle

\section{Introduction}
One generally believed feature of quantum gravity is the existence
of a minimal observable length\cite{Gross87ar,Maggiore93rv}. Recent
development of loop quantum gravity has also greatly strengthen such
beliefs by manifestly showing the discreteness of area and volume
spectra\cite{Rovelli94ge}. At the same time such effects have also
invoked many investigations and debates on the fate of Lorentz
symmetry at Planck scale\cite{Amelino00ge}. One reason is that such
effects seemly lead to a paradox due to the apparent confliction
between the existence of a minimum length and Lorentz symmetry,
which in principle may contract any object to arbitrary small size
by Lorentz boost. Nowadays one intriguing approach dubbed as doubly
special relativity (DSR) is proposed to solve this paradox, for
details see recent review\cite{Amelino03uc} and references therein.
In particular, a general formalism of modifying special relativity
has been proposed in \cite{Magueijo01cr} to preserve the relativity
of inertial frames, while at the same time keep a physical energy
such as Planck energy as an invariant. This is accomplished by a
non-linear action of the Lorentz transformation in momentum space.
This formalism also points to the possibility that the usual
relation between the energy and momentum in special relativity may
be modified at Planck scale, conventionally named as modified
dispersion relations (MDR). Such relations can also be derived in
the study of semi-classical limit of loop quantum
gravity\cite{Gambin98it,Smolin05cz}.

The modification of energy-momentum relations and its implications
have been greatly investigated by many
theorists\cite{Unruh94je,Colladay98fq,Coleman98ti,Amelino00zs,Foster05fr}.
It may be responsible for some peculiar phenomena in experiments and
astronomic observations, such as the threshold anomalies of ultra
high energy cosmic ray and Gamma ray burst. Such modifications may
further lead to some predictions which can be falsified in planned
experiments. Among these are the energy dependence of the speed of
light and the helicity independence of dispersion relations,
observable in the coming AUGER and GLAST
experiments\cite{Smolin05cz}. Moreover, a modified dispersion
relations may present alternatives to inflationary
cosmology\cite{Moffat92ud}, and this is testable in the future
measurement of CMB spectrum.

In this paper we intend to study the impact of modified dispersion
relations on black hole physics. We first present a modified
dispersion relation advocated by doubly special relativity, and
then show that this modified relation may contribute corrections
to the temperature of black holes as well as the entropy. We find
the entropy has a logarithmic correction while the temperature is
bounded with a finite value as the mass of black holes approaches
to the Planck scale such that black holes will finally stop
radiating, in contrast to the ordinary picture where the
temperature can be arbitrary high as the mass approaches to zero
and finally divergent when black holes fully evaporate. A
comparison with effects due to the generalized uncertainty
principle is also discussed.

\section{Modified dispersion relations }

As pointed out in \cite{Magueijo01cr}, in a DSR framework the
modified dispersion relation may be written as \f
E^2f_1^2(E;\eta)-P^2f_2^2(E;\eta)=m_0^2,\ff where $f_1$ and $f_2$
are two functions of energy from which a specific formulation of
boost generator can be defined. In this paper we adopt a modified
dispersion relation (MDR) by taking $f_1^2=[1-\eta(l_pE)^n]$ and
$f_2^2=1$, such that \f E^2={p^2+m_0^2\over [1-\eta(l_pE)^n]},\ff
where Planck length $l_p\equiv \sqrt{8\pi G}\equiv 1/M_p$ and
$\eta$ is a dimensionless parameter. If $l_pE\ll 1$, this modified
dispersion relation goes back to the ordinary one \f
E^2=p^2+m_0^2+\eta (l_pE)^n(p^2+m_0^2+...).\ff

However, when $l_pE\sim 1$ the relation changes greatly and we
need treat it non-perturbatively.  For convenience, we take $n=2$
and $\eta =1$, \f l_p^2E^4-E^2+(p^2+m_0^2)=0.\label{n2}\ff

This gives a relation as \f E^2={1\over
2l_p^2}\left[1-\sqrt{1-4l_p^2(p^2+m_0^2)}\right]\label{mep}.\ff

Due to the appearance of square root in above equation, we notice
that both the momentum and the static mass of a single particle are
bounded by $m_0\leq M_p/2$ and $p^2\leq (M_p^2/4-m_0^2)$, in
particular for massless particles the limiting momentum is $M_p/2$.
The existence of a maximum momentum reflects the feature that there
is a minimal observable length in quantum gravity, and more delicate
analysis can be found in \cite{Gross87ar,Maggiore93rv,Adler01vs},
where generalized uncertainty principle is employed, while in the
context of DSR our argument here can be considered as a result of
the non-linear Lorentz transformation in momentum
space\cite{Amelino03uc,Magueijo01cr}. From (\ref{mep}) it is also
easy to see that a single particle has a maximum energy
$E_{max}=M_p/\sqrt{2}$.

\section{Linking modified dispersion relations to black hole physics}

Now we consider the impact of this MDR on Schwarzschild black holes.
The effect that the existence of a minimum length can prevent black
holes from total evaporation has been investigated in
\cite{Adler01vs}, where the generalized uncertainty principle (GUP)
plays an essential role. Here we will consider a modified dispersion
relation rather than GUP. First we identify above quantities $E$ and
$p$ as the energy and momentum of photons emitted from the black
hole respectively, of course for photons $m_0$ is set to zero. Then
we adopt the argument presented in\cite{Adler01vs} that the
characteristic temperature of this black hole is supposed to be
proportional to the photon energy $E$,namely $E=T$. On the other
hand, we apply the ordinary uncertainty relation to photons in the
vicinity of black hole horizons. As pointed out in
\cite{AH,Adler01vs}, for these photons there is an intrinsic
uncertainty about the Schwarzschild radius $R$. \f p\sim \delta
p\sim {1\over\delta x}\sim {1\over 4\pi
R},\ff where %$R$ is the radius of Schwarzschild black hole and
a ``calibration factor'' $4\pi$ is introduced. Using the fact that
$R=2MG\sim M/ 4\pi M_p^2$\footnote{We assume that the modified
dispersion relation will not change this relation which is
consistent with the result in \cite{Magueijo02xx}, where the
impact of doubly special relativity on gravity is investigated.
Also see \cite{LLZ} for recent work on the thermodynamics of
modified black holes from gravity's rainbow. }, we obtain the
temperature of Schwarzschild black holes has the form \f T= \left[
{M_p^2\over 2}\left(1-\sqrt{1-{4M_p^2\over
M^2}}\right)\right]^{1/2}.\label{tem}\ff

This requires that the mass of black holes $M\geq 2M_p$, and
correspondingly the temperature  $T\leq M_p/ \sqrt{2}$ \footnote{
It seems possible to relate this formalism to the Immirzi
parameter $\gamma$ in loop quantum gravity where
$A_{min}=\sqrt{3}/2 \gamma l_p^2$, since in dispersion relations
we can freely add some parameter $\eta$ in the term $(l_pE)^2$ in
(\ref{n2}) as we pointed out. Then setting these two minimum area
equal leads to a relation $\gamma=2\eta/(\sqrt{3}\pi)$.}. For
large black holes with $M\gg 2M_p$, it goes back to the ordinary
form $T= M_p^2/ M$.%\f T= {M_p^2\over M}.\ff

Next we consider the possible correction to the entropy of black
holes due to the modification of the temperature, assuming the
first thermodynamical law still exactly holds even for small black
holes where the quantum effect of gravity may play an essential
role. \footnote{ We make this assumption based on the belief that
the thermodynamical laws of black holes can still be captured even
in the quantum theory of gravity. This belief has been supported
by recent work in isolated horizon programm as well as the
stretched horizon programm, where the first law of black hole
thermodynamics can still be established in a quasi-local fashion.
In the context of DSR, which may be viewed as the semi-classical
effect of quantum gravity, we assume the modification of
dispersion relations of particle would not change this picture.
However, before proceed we cautiously point out that this exact
form of the first law may not be generally true but receive a
small modification, or even fails for instance in the
Einstein-Aether theory as investigated in \cite{Foster05fr}. }.
Plugging the temperature into $dM=TdS$, we have, \f dM= \left[
{M_p^2\over 2}\left( 1-\sqrt{1-{4M_p^2\over
M^2}}\right)\right]^{1/2}dS.\ff Thus, the entropy can be
calculated from the integration, \f S={1\over
2\sqrt{G}}\int_{A_{min}}^A(A-\sqrt{A^2-8GA})^{-1/2}dA
,\label{en}\ff where $A_{min}=8G\sim {l_p^2/ \pi}$ is the cutoff
corresponding to a black hole with minimum mass $M=2M_p$. Define
$t=\sqrt{1-8G/A}$, (\ref{en}) can be integrated out \fa S &=&
{1\over \sqrt{2}}\left[ \left. 2(1+t)^{-1/2}\right\vert
_{t_{min}}^{t}+\left. \frac{\sqrt{1+t}}{1-t}\right\vert
_{t_{min}}^{t}-\right. \nonumber\\ && \left.
\frac{1}{\sqrt{2}}\left.
\ln(\frac{1+t}{1-t})\right\vert _{\sqrt{1+t_{min}}/\sqrt{2}}^{\sqrt{1+t}%
/\sqrt{2}}\right]+S_{min}.\label{er} \ffa

where $S_{min}=A_{min}/4G$ is a constant term  such that for
minimum black holes the familiar Bekenstein-Hawking entropy
formula still holds. When $A\gg 8G$, it becomes \f S= {1\over
\sqrt{2}}\left[ {A\over 8G}(1+t)^{3/2}-{1\over \sqrt{2}}ln[{A\over
8G}(1+t)]\right].\ff It is obvious that for large black holes,
namely $t\rightarrow 1$, this gives the familiar formula \f
S={A\over 4G}-{1\over 2}ln{A\over 4G}+...\label{ec1}\ff Therefore,
in this case we find modified dispersion relations contributes a
logarithmic correction to black hole entropy\footnote{ The factor
$-1/2$ in the logarithmic term happens to be the same as the one
appearing in \cite{Meissner04ju} where this factor is rigorously
fixed in the context of loop quantum gravity. However, we stress
that at semi-classical level as discussed in our paper this factor
can not be fixed uniquely, but depend on the value of $\eta$ which
has been set as unit in our paper. A straightforward calculation
shows that in general the factor would be $-\eta/2$. If we insist
that the minimum area in our paper is the same as that in loop
quantum gravity as we suggested in above footnote, then the factor
turns out to be $-\sqrt{3}\pi\gamma/4$, rather than $-1/2$.
Reversely, if we insist to set the factor exactly to be $-1/2$,
namely $\eta =1$, then from this approach we have a Immirzi
parameter $\gamma\simeq 0.37$, different from the result in
\cite{Meissner04ju}. This discrepancy can be understood as this
specific MDR we proposed is only a coarse grained model at
semi-classical limit of quantum gravity.}.

The entropy correction can also be evaluated using the scheme
proposed in \cite{Amelino04xx}, where the Bekenstein entropy
assumption is applied to determine the minimum increase of horizon
area when a black hole absorbs a classical particle with energy
$\epsilon$ and size $\delta x$. Explicitly, the assumption is \f
{\delta A_{min}\over 4G}\geq 2\pi \epsilon \delta x.\ff On the other
hand, from (\ref{mep}) we may obtain a uncertainty relation between
the energy and momentum of a single particle as \f \delta E \sim
{\delta p\over (1-2l_p^2p^2)}.\label{ur}\ff Identifying
$\epsilon\sim \delta E$ and $\delta p\sim 1/\delta x\sim 1/(4\pi
R)$, we may have \f \delta A_{min}\cong 16\pi Gln2 \epsilon \delta
x={4Gln2\over 1- l_p^2/ 2\pi A},\ff where a calibration factor
$2ln2$ is introduced. Thus \f {dS\over dA}\cong {\delta S\over
\delta A_{min}}={1\over 4G}(1-{l_p^2\over 2\pi A}),\ff and
consequently \f S={A\over 4G}-ln\left({A\over
4G}\right)\label{ec2}.\ff Comparing with (\ref{ec1}) we find the
answer is almost the same but the factor in logarithmic term is
different, which results from the fact that some approximations for
large black holes have been taken into account during
 the derivation of (\ref{ec2}), for instance in (\ref{ur}). In this sense
we argue that we find a more precise way to obtain corrections to
black hole entropy. This can also be understood from the fact that
in our case the exact expression for entropy corresponding to the
temperature (\ref{tem}) can be obtained, as shown in (\ref{er}), in
contrast to \cite{Amelino04xx} where such exact expressions are not
available. Moreover, it is worthwhile to point out that in those
papers the entropy is obtained approximately at first, and then the
temperature is derived with the use of the first thermodynamical
law. With no surprise, such a logic gives rise to a different result
for temperature from (\ref{tem}), but approximately equal at large
black hole limit. From this point of view, our results are more
general. It is also this advantage that make them applicable to
investigate the fate of black holes at the late stage of radiation.
We briefly present our analysis below, as a similar discussion has
appeared in \cite{Adler01vs}.

Thanks to the Stefan-Boltzmann law\footnote{We neglect the
possible modification of this law due to the existence of a cutoff
for the thermal spectrum here, which may change the value of
Stefan-Boltzmann constant but not affect our conjecture about the
black hole remnant in the paper.}, the evaporation rate of black
holes can be estimated by, \f {dx\over dt} ={-1\over t_{f}}\left(
x-\sqrt{x^2-4}\right)^2,\label{rad}\ff where $x=M/M_p$ and
$t_f=16\pi/(\sigma M_p)$. The solution to this equation reads as
\f t=t_c-{t_f\over 24}\left[ x^3+(x^2-4)^{3/2}-6x\right],\ff where
$t_c$ is an integral constant. From above equation we find that
$dx/dt=-4/t_f$ is a finite number at the end $x=2$ rather than
infinity in ordinary case. As a matter of fact when the size of
black holes approaches the Planck scale, they will cease radiation
although the temperature reaches a maximum. This can be seen from
the behavior of the heat capacity. From Eq.(\ref{tem}), we obtain
\begin{eqnarray}
C=\frac{dM}{dT}=-\frac{M^3T}{M_p^4}\cdot
\left(1-\frac{4M_p^2}{M^2}\right)^{1/2}.
\end{eqnarray}
It is interesting that the heat capacity becomes vanishing when
the black hole mass approaches a nonzero scale, $M=2M_p$. This
maybe implies the ground state of the black hole. As an analogy,
let us look at a system consisting of the harmonic oscillators,
the heat capacity is vanishing when the system is in the ground
state. This is because the  zero energy is independent of the
temperature, $\partial E_0/\partial T=0$. This phenomena provides
a mechanism to take black hole remnants as a natural candidate for
cold dark matter due to their weakly interacting
features\cite{Barrow92hq}.

At the end of this section we point out that the effect of
generalized uncertainty principle will not change our conclusions
but provide modifications. The well-known uncertainty relation can
be generalized as \f \delta x \delta p\geq 1 + l_p^2\delta p^2.\ff
Thus \f \delta p={\delta x\over 2l_p^2}\left(1\mp
\sqrt{1-4l_p^2/\delta x^2}\right).\ff Set $p\sim \delta p$ and
plug it into equation (\ref{mep}), we obtain \f T\sim \left[
{M_p^2\over
2}\left(1-\sqrt{5-2x(x-\sqrt{x^2-4})}\right)\right]^{1/2},\ff
where $x=M/M_p\geq 5/2$. Thus the black hole ceases radiation as
it reaches this minimum value at Planck scale.

\section{Concluding remarks}
In this paper we have shown that the modified dispersion relations
have important impacts on black hole physics at high energy level.
First MDR contributes a correction to the temperature of black
holes and provides an effective cutoff such that a upper limit
will be reached as the area of black hole horizon takes the
minimum value. Correspondingly, the black hole entropy is
corrected with a logarithmic term. Secondly, MDR provides a
plausible mechanism to prevent black holes from fully evaporating
and the remnant can be treated as a candidate for cold dark
matter.

Through the paper we only investigate a special case in doubly
special relativity with specified functions of $f_1(E;\eta)$ and
$f_2(E;\eta)$, but it is obviously possible to extend our
discussion to other general cases. For instance, we may take $n=1$
such that $f_1^2=[1-\eta(l_pE)]$, a parallel analysis can be done
and the entropy of black holes is expected to receive a correction
proportional to the square root of the area. Among all the
possible modified dispersion relations which is the proper one
should await further tests in experiments.

It is interesting to notice that both generalized uncertainty
relations and modified dispersion relations may be rooted at the
algebraic structure of commutators among position and momentum
variables. More deep relations between of them and implications to
quantum gravity phenomenology are under investigation.

\section*{Acknowledgement}
Y.L. would like to thank Prof. Hoi-Lai Yu for his hospitality during
the visit at Institute of Physics, Academia Sinica in Taiwan, where
this work is completed. This work is partly supported by NSFC
(No.10405027, 10205002, 10505011) and SRF for ROCS, SEM.


\begin{thebibliography}{9}                                                                                                %

%\cite{Gross:1987ar}
\bibitem{Gross87ar}
  D.~J.~Gross and P.~F.~Mende,
  %``String Theory Beyond The Planck Scale,''
  Nucl.\ Phys.\ B {\bf 303}, 407 (1988).
  %%CITATION = NUPHA,B303,407;%%


%\cite{Maggiore:1993rv}
\bibitem{Maggiore93rv}
  M.~Maggiore,
  % ``A Generalized uncertainty principle in quantum gravity,''
  %
  Phys.\ Lett.\ B {\bf 304}, 65 (1993).
  %[arXiv:hep-th/9301067].

  %%CITATION = HEP-TH 9301067;%%
%\cite{Rovelli:1994ge}
\bibitem{Rovelli94ge}
  C.~Rovelli and L.~Smolin,
  %``Discreteness of area and volume in quantum gravity,''
  Nucl.\ Phys.\ B {\bf 442}, 593 (1995)
  [Erratum-ibid.\ B {\bf 456}, 753 (1995)],
 % [arXiv:gr-qc/9411005].
  %%CITATION = GR-QC 9411005;%%
%\cite{Ashtekar:1996eg}
%\bibitem{Ashtekar96eg}
  A.~Ashtekar and J.~Lewandowski,
  %``Quantum theory of geometry. I: Area operators,''
  Class.\ Quant.\ Grav.\  {\bf 14}, A55 (1997),
  %[arXiv:gr-qc/9602046].
  %%CITATION = GR-QC 9602046;%%
%\cite{Ashtekar:1997fb}
%\bibitem{Ashtekar:1997fb}
 % A.~Ashtekar and J.~Lewandowski,
  %``Quantum theory of geometry. II: Volume operators,''
   Adv.\ Theor.\ Math.\ Phys.\  {\bf 1}, 388 (1998),
 % [arXiv:gr-qc/9711031].
  %%CITATION = GR-QC 9711031;%%
  %\cite{Brunnemann:2004xi}
%\bibitem{Brunnemann04xi}
  J.~Brunnemann and T.~Thiemann,
  %``Simplification of the spectral analysis of the volume operator in loop
  %quantum gravity,''
  Class.\ Quant.\ Grav.\ {\bf 23}, 1289 (2006).
 % arXiv:gr-qc/0405060.
  %%CITATION = GR-QC 0405060;%%

%\cite{Amelino-Camelia:2000ge}
\bibitem{Amelino00ge}
  G.~Amelino-Camelia,
   %``Testable scenario for relativity with minimum-length,''
  %
  Phys.\ Lett.\ B {\bf 510}, 255 (2001),
 % [arXiv:hep-th/0012238].
  %%CITATION = HEP-TH 0012238;%%
%\cite{Amelino-Camelia:2000mn}
%\bibitem{Amelino00mn}
  G.~Amelino-Camelia,
  % ``Relativity in space-times with short-distance structure governed by an
  % observer-independent (Planckian) length scale,''
  %
  Int.\ J.\ Mod.\ Phys.\ D {\bf 11}, 35 (2002),
 % [arXiv:gr-qc/0012051].
  %%CITATION = GR-QC 0012051;%%
%\cite{Amelino-Camelia:2003ex}
%\bibitem{Amelino03ex}
  G.~Amelino-Camelia, J.~Kowalski-Glikman, G.~Mandanici and A.~Procaccini,
  % ``Phenomenology of doubly special relativity,''
  Int.\ J.\ Mod.\ Phys.\ A {\bf 20}, 6007 (2005),
 % arXiv:gr-qc/0312124.
  %%CITATION = GR-QC 0312124;%%
%\cite{Jacobson:2001tu}
%\bibitem{Jacobson01tu}
  T.~Jacobson, S.~Liberati and D.~Mattingly,
 %  ``TeV astrophysics constraints on Planck scale Lorentz violation,''
  %
  Phys.\ Rev.\ D {\bf 66}, 081302 (2002).
 % [arXiv:hep-ph/0112207].
  %%CITATION = HEP-PH 0112207;%%
%\cite{Amelino03uc}
\bibitem{Amelino03uc}
  G.~Amelino-Camelia,
   ``The three perspectives on the quantum-gravity problem and their
   implications for the fate of Lorentz symmetry,''
  %
  arXiv:gr-qc/0309054.
  %%CITATION = GR-QC 0309054;%%

%\cite{Magueijo:2001cr}
\bibitem{Magueijo01cr}
  J.~Magueijo and L.~Smolin,
  %``Lorentz invariance with an invariant energy scale,''
  Phys.\ Rev.\ Lett.\  {\bf 88}, 190403 (2002),
 % [arXiv:hep-th/0112090].
  %%CITATION = HEP-TH 0112090;%%
%\cite{Magueijo:2002am}
%\bibitem{Magueijo02am}
  J.~Magueijo and L.~Smolin,
  %``Generalized Lorentz invariance with an invariant energy scale,''
  Phys.\ Rev.\ D {\bf 67}, 044017 (2003).
 % [arXiv:gr-qc/0207085].
  %%CITATION = GR-QC 0207085;%%

%\cite{Gambini:1998it}
\bibitem{Gambin98it}
  R.~Gambini and J.~Pullin,
  % ``Nonstandard optics from quantum spacetime,''
  %
  Phys.\ Rev.\ D {\bf 59}, 124021 (1999),
 % [arXiv:gr-qc/9809038].
  %%CITATION = GR-QC 9809038;%%
%\cite{Alfaro:2001rb}
%\bibitem{Alfaro01rb}
  J.~Alfaro, H.~A.~Morales-Tecotl and L.~F.~Urrutia,
  % ``Loop quantum gravity and light propagation,''
  %
  Phys.\ Rev.\ D {\bf 65}, 103509 (2002),
  %[arXiv:hep-th/0108061].
  %%CITATION = HEP-TH 0108061;%%
%\cite{Smolin:2002sz}
%\bibitem{Smolin02sz}
  L.~Smolin,
  %``Quantum gravity with a positive cosmological constant,''
  arXiv:hep-th/0209079,
  %%CITATION = HEP-TH 0209079;%%
%\cite{Sahlmann:2002qk}
%\bibitem{Sahlmann02qk}
  H.~Sahlmann and T.~Thiemann,
  %``Towards the QFT on curved spacetime limit of QGR. II: A concrete
  %implementation,''
  Class.\ Quant.\ Grav. {\bf 23}, 909 (2006).
  % arXiv:gr-qc/0207031.
  %%CITATION = GR-QC 0207031;%%
%\cite{Smolin:2005cz}
\bibitem{Smolin05cz}
  L.~Smolin,
  ``Falsifiable predictions from semiclassical quantum gravity,''
  arXiv:hep-th/0501091.
  %%CITATION = HEP-TH 0501091;%%

%\cite{Unruh:1994je}
\bibitem{Unruh94je}
  W.~G.~Unruh,
  %``Sonic analog of black holes and the effects of high frequencies on black
  %hole evaporation,''
  Phys.\ Rev.\ D {\bf 51}, 2827 (1995).
  %%CITATION = PHRVA,D51,2827;%%

%\cite{Colladay:1998fq}
\bibitem{Colladay98fq}
  D.~Colladay and V.~A.~Kostelecky,
  %``Lorentz-violating extension of the standard model,''
  Phys.\ Rev.\ D {\bf 58}, 116002 (1998).
 % [arXiv:hep-ph/9809521].
  %%CITATION = HEP-PH 9809521;%%

 %\cite{Coleman:1998ti}
\bibitem{Coleman98ti}
  S.~R.~Coleman and S.~L.~Glashow,
  %``High-energy tests of Lorentz invariance,''
  Phys.\ Rev.\ D {\bf 59}, 116008 (1999).
 % [arXiv:hep-ph/9812418].
  %%CITATION = HEP-PH 9812418;%%

%\cite{Amelino-Camelia:2000zs}
\bibitem{Amelino00zs}
  G.~Amelino-Camelia and T.~Piran,
  % ``Planck-scale deformation of Lorentz symmetry as a solution to the UHECR
  % and the TeV-gamma paradoxes,''
  %
  Phys.\ Rev.\ D {\bf 64}, 036005 (2001),
 % [arXiv:astro-ph/0008107].
  %%CITATION = ASTRO-PH 0008107;%%
%\cite{Myers:2003fd}
%\bibitem{Myers03fd}
R.~C.~Myers and M.~Pospelov,
%``Experimental challenges for quantum gravity,''
Phys.\ Rev.\ Lett.\  {\bf 90}, 211601 (2003),
%[arXiv:hep-ph/0301124].
%%CITATION = HEP-PH 0301124;%%
%\cite{Jacobson:2003bn}
%\bibitem{Jacobson03bn}
  T.~A.~Jacobson, S.~Liberati, D.~Mattingly and F.~W.~Stecker,
  % ``New limits on Planck scale Lorentz violation in QED,''
  %
  Phys.\ Rev.\ Lett.\  {\bf 93}, 021101 (2004),
 % [arXiv:astro-ph/0309681].
 S.~Corley and T.~Jacobson,
  %``Black hole lasers,''
  Phys.\ Rev.\ D {\bf 59}, 124011 (1999),
%  [arXiv:hep-th/9806203].
  %%CITATION = HEP-TH 9806203;%%
%\cite{Hossenfelder:2003jz}
%\bibitem{Hossenfelder03jz}
  S.~Hossenfelder \textit{et al.}, % M.~Bleicher, S.~Hofmann, J.~Ruppert, S.~Scherer and
%H.~Stoecker,
  %``Collider signatures in the Planck regime,''
  Phys.\ Lett.\ B {\bf 575}, 85 (2003),
 % [arXiv:hep-th/0305262].
  %%CITATION = HEP-TH 0305262;%%
%\cite{Hossenfelder:2005ed}
%\bibitem{Hossenfelder:2005ed}
  S.~Hossenfelder,
  %``Self-consistency in theories with a minimal length,''
  Class.\ Quantum\ Grav.\  {\bf 23}, 1815 (2006).
 % arXiv:hep-th/0510245.
  %%CITATION = HEP-TH 0510245;%%

  %%CITATION = ASTRO-PH 0309681;%%

%\cite{Foster:2005fr}
\bibitem{Foster05fr}
  B.~Z.~Foster,
  %``Noether charges and black hole mechanics in Einstein-aether theory,''
  Phys.\ Rev.\ D {\bf 73}, 024005 (2006).
 % [arXiv:gr-qc/0509121].
  %%CITATION = GR-QC 0509121;%%

%\cite{Moffat:1992ud}
\bibitem{Moffat92ud}
  J.~W.~Moffat,
  % ``Superluminary universe: A Possible solution to the initial value problem in
  % cosmology,''
  %
  Int.\ J.\ Mod.\ Phys.\ D {\bf 2}, 351 (1993),
  %[arXiv:gr-qc/9211020].
  %%CITATION = GR-QC 9211020;%%
%\cite{Albrecht:1998ir}
%\bibitem{Albrecht98ir}
  A.~Albrecht and J.~Magueijo,
  % ``A time varying speed of light as a solution to cosmological puzzles,''
  %
  Phys.\ Rev.\ D {\bf 59}, 043516 (1999).
 % [arXiv:astro-ph/9811018].
  %%CITATION = ASTRO-PH 9811018;%%


%\cite{Adler:2001vs}
\bibitem{Adler01vs}
  R.~J.~Adler, P.~Chen and D.~I.~Santiago,
  %``The generalized uncertainty principle and black hole remnants,''
  Gen.\ Rel.\ Grav.\  {\bf 33}, 2101 (2001),
 % [arXiv:gr-qc/0106080].
  %%CITATION = GR-QC 0106080;%%
%\cite{Chen:2002tu}
%\bibitem{Chen02tu}
  P.~Chen and R.~J.~Adler,
  %``Black hole remnants and dark matter,''
  Nucl.\ Phys.\ Proc.\ Suppl.\  {\bf 124}, 103 (2003).
 % [arXiv:gr-qc/0205106].
  %%CITATION = GR-QC 0205106;%%

\bibitem{AH}
  R.~J.~Adler and T.~K.~Das, Phys.\ Rev.\ D {\bf 14}, 2474 (1976).
  R.~S.~Hanni and R. ~Ruffini, ``Lines of force of a point charge
  near a Schwarzschild black hole", in {\it Black Holes},
  eds.~C.~DeWitt and B.~S.~DeWitt (Gordon Breach,1973).

%\cite{Amelino-Camelia:2004xx}
\bibitem{Amelino04xx}
  G.~Amelino-Camelia, M.~Arzano and A.~Procaccini,
  %``Severe constraints on loop-quantum-gravity energy-momentum dispersion
  %relation from black-hole area-entropy law,''
  Phys.\ Rev.\ D {\bf 70}, 107501 (2004),
 % [arXiv:gr-qc/0405084].
  %%CITATION = GR-QC 0405084;%%
%\cite{Amelino-Camelia:2005ik}
%\bibitem{Amelino05ik}
  G.~Amelino-Camelia, M.~Arzano, Y.~Ling and G.~Mandanici,
  %``Black-hole thermodynamics with modified dispersion relations and
  %generalized uncertainty principles,''
   Class.Quantum Grav. 23, 2585 (2006),
  % arXiv:gr-qc/0506110.
  %%CITATION = GR-QC 0506110;%%
%\cite{Medved04yu}
%\bibitem{Medved04yu}
  A.~J.~M.~Medved and E.~C.~Vagenas,
  %``When conceptual worlds collide: The GUP and the BH entropy,''
  Phys.\ Rev.\ D {\bf 70}, 124021 (2004).
 % [arXiv:hep-th/0411022].
  %%CITATION = HEP-TH 0411022;%%

%\cite{Setare:2004sr}
%\bibitem{Setare:2004sr}
%  M.~R.~Setare,
  %``Corrections to the Cardy-Verlinde formula from the generalized uncertainty
  %principle,''
%  Phys.\ Rev.\ D {\bf 70}, 087501 (2004)
 % [arXiv:hep-th/0410044].
  %%CITATION = HEP-TH 0410044;%%

%\cite{Barrow:1992hq}
\bibitem{Barrow92hq}
 J.H.MacGibbon,
Nature {\bf 329}, 308 (1987).
%%CITATION = NATUA,329,308;%%
 J.~D.~Barrow, E.~J.~Copeland and A.~R.~Liddle,
   %``The Cosmology of black hole relics,''
  %
  Phys.\ Rev.\ D {\bf 46}, 645 (1992).
  %%CITATION = PHRVA,D46,645;%%
  B.~J.~Carr, J.~H.~Gilbert and J.~E.~Lidsey,
 %  ``Black hole relics and inflation: Limits on blue perturbation spectra,''
  %
  Phys.\ Rev.\ D {\bf 50}, 4853 (1994).
%  [arXiv:astro-ph/9405027].
  %%CITATION = ASTRO-PH 9405027;%%

%\cite{Magueijo:2002xx}
\bibitem{Magueijo02xx}
  J.~Magueijo and L.~Smolin,
  % ``Gravity's Rainbow,''
  %
  Class.\ Quant.\ Grav.\  {\bf 21}, 1725 (2004).
 % [arXiv:gr-qc/0305055].
  %%CITATION = GR-QC 0305055;%%

\bibitem{LLZ}
Y. Ling, X.Li and H.Zhang, ``Thermodynamics of modified black
holes from gravity's rainbow'', [arXiv:gr-qc/0512084].

%\cite{Meissner:2004ju}
\bibitem{Meissner04ju}
  K.~A.~Meissner,
  %``Black hole entropy in loop quantum gravity,''
  Class.\ Quant.\ Grav.\  {\bf 21}, 5245 (2004).
%  [arXiv:gr-qc/0407052].
  %%CITATION = GR-QC 0407052;%%

\end{thebibliography}
\end{document}